# Protocol Optimization for Functional Cardiac CT Imaging Using Noise Emulation in the Raw Data Domain

Zhye Yin[1*], Pengwei Wu[2*], Ashish Manohar[3], Elliot R. McVeigh[4], and Jed D. Pack[2]

[1]GE HealthCare, Waukesha, WI, USA

[2]GE Research Healthcare, Niskayuna, NY, USA

[3]Dept. of Medicine, Stanford University, Palo Alto, CA, USA

[4]Dept. of Bioengineering, Medicine, Radiology at University of California San Diego, San Diego, CA, USA

*These two authors contributed equally

`

**Corresponding Author:**

Jed D. Pack, Ph.D.

1 Research Circle

Niskayuna, NY 12309, USA

Phone: +1 (518) 4109503

Email: jed.pack@ge.com

## Abstract

**Background:** Four-dimensional (4D) wide coverage computed tomography (CT) is an effective imaging modality for measuring the mechanical function of the myocardium. However, repeated CT measurement across a number of heartbeats is still a concern.

**Purpose:** A projection-domain noise emulation method is presented to generate accurate low-dose (mA modulated) 4D cardiac CT scans from high-dose scans, enabling protocol optimization to deliver sufficient image quality for functional cardiac analysis while using a dose level that is as low as reasonably achievable (ALARA).

**Methods:** Given a targeted low-dose mA modulation curve, the proposed noise emulation method injects both quantum and electronic noise of proper magnitude and correlation to the high-dose data in projection domain. A spatially varying (i.e., channel-dependent) detector gain term as well as its calibration method were proposed to further improve the noise emulation accuracy. To determine the ALARA dose threshold, a straightforward projection domain image quality (IQ) metric was proposed that is based on the number of projection rays that do not fall under the non-linear region of the detector response. Experiments were performed to validate the noise emulation method with both phantom and clinical data in terms of visual similarity, contrast-to-noise ratio (CNR), and noise-power spectrum (NPS).

**Results:** For both phantom and clinical data, the low-dose emulated images exhibited similar noise magnitude (CNR difference within 2%), artifacts, and texture to that of the real low-dose images. The proposed channel-dependent detector gain term resulted in additional increase in emulation accuracy. Using the proposed IQ metric, recommended kVp and mA settings were calculated for low dose 4D Cardiac CT acquisitions for patients of different sizes.

**Conclusions:** A detailed method to estimate system-dependent parameters for a raw-data based low dose emulation framework was described. The method produced realistic noise levels, artifacts, and texture with phantom and clinical studies. The proposed low-dose emulation method can be used to prospectively select patient-specific minimal-dose protocols for functional cardiac CT.

# I. Introduction

Wide coverage computed tomography (CT) is an effective imaging modality for measuring the mechanical function of the myocardium. The coverage of 16 cm eliminates the needs of multiple table positions and provides high fidelity 4D images of a beating heart from a single heartbeat acquisition. Characterizing dyssynchronous LV wall motion available from the 4D CT images has recently been shown to predict response to Cardiac Resynchronization Therapy (CRT),[1] a medical procedure aimed at correcting mechanical dyssynchrony.[2–5] Also, myocardial perfusion measurements require multiple images across a number of heartbeats. Hence, the X-ray dose from these 4D acquisitions is still a concern.[6] There is a clear compromise between radiation dose and the resulting image quality for downstream 4D CT analysis. The purpose of this work is to optimize acquisition parameters, such as X-ray tube current (mA) and tube voltage (kVp), to deliver sufficient image quality for functional cardiac analysis techniques while using a dose that is as low as reasonably achievable (ALARA).

Ideally, to investigate the implications of low dose on cardiac function analysis, multiple acquisitions would be obtained in the same subject during precisely the same motions; however, this is not possible for patient studies. Alternatively, inserting synthetic noise to existing clinical data sets has been explored.[7–21] Synthetic noise insertion is referred to as low dose emulation (LDE) in this work. Common LDE approaches can be classified into two broad categories: image-based and raw data-based. Image based methods add noise into reconstructed CT images without accessing projection domain raw data.[17–21] These approaches are useful when the raw data are not available. However, non-local noise properties make it quite difficult to accurately insert noise in the image domain.[8] The raw data-based approach simulates lower-dose scans by adding additional noise in the projection domain.[7–16] Raw data-based approaches are often more realistic especially at extremely low dose levels because: (i) the negative log and associated pre-processing steps create nonlinearity at low flux levels which is not easily reproduced with image domain approaches;[22,23] (ii) the noise is more accurately modeled in the projection domain (for example, as a combination of Poisson-distributed quantum noise and Gaussian-distributed electronic noise).

Considering these advantages, a new raw data-based / projection domain LDE approach is proposed in this work. Prior approaches in the literature typically assume that the noise distribution of the quantum and electronic noise at each channel is the same, given the same mean signal value and signal spectrum. While this assumption is reasonable, it is not perfectly accurate considering that the conversion factor / gain from photons to electrons is in fact channel dependent. Therefore, the previously proposed LDE methods can be improved by modeling and calibrating channel-specific properties of the detector as done in this work. In addition, our LDE approach includes modeling of the bowtie filters, beam hardening (BH) effects from the patients, nonlinearity in detector response at extremely low flux conditions, and the temporal correlation of electronic noise.

Our goal is to provide a critical tube current for a given patient at a given voltage, below which an acquisition will fail to meet a desired image quality (IQ) requirement. In this work, instead of using an image-based IQ metric, we propose a new raw data-based IQ metric to determine the minimum flux an acquisition can tolerate while still generating usable information for a given application. This IQ metric is useful for complex applications such as measuring myocardial function where a universal image quality requirement is difficult to formulate. Additionally, a projection domain IQ metric is much more straightforward to calculate without the need for actual reconstructions. This is especially useful for 4D CT applications where many reconstructions (e.g., multiple phases in 4D cardiac CT) are required for each tube current and voltage level.

We first describe our new projection-domain LDE approach in section II.A-II.C. The protocol optimization method utilizing the proposed projection domain IQ metric is described in section II.D. Section III presents the validation of the LDE approach, and the results of critical mA-based protocol optimization method with phantom and clinical studies. Finally, some concluding remarks are drawn in section IV. The novel contributions of this work are as follows: (i) while applicable to other CT applications, the proposed method is heavily focused on 4D functional cardiac CT applications such as ejection fraction or myocardial strain, which typically can be performed with low mA and mA modulation, (ii) a new LDE and calibration approach is proposed which characterizes spatially-varying response functions for each detector channel; (iii) a new projection domain IQ metric is proposed for efficient scan protocol optimization; (iv) recommended kVp and mA settings were reported for low dose 4D Cardiac CT acquisitions for patients of different sizes.

## II. Methods

Figure 1 illustrates our LDE strategy and the general workflow for selecting the minimum-dose protocol for CT scans for cardiac patients. Given a non-mA-modulated full dose (FD) scan covering a full cardiac cycle [example mAs curve shown in Fig. 1(a)], the proposed LDE method can be used to generate emulated low dose (LD) projections given a targeted mAs modulation curve [example mAs shown in Fig. 1(b), where high mAs / dose is only prescribed for the mid-diastole phase].

The protocol selection step (blue boxes in Fig. 1) is patient-specific and can be done in either projection or image domain, or both. As mentioned above, projection domain IQ metrics are much easier to calculate as they do not require (multi-phase) reconstructions for each targeted mAs modulation curve. Image domain metrics on the other hand could account for the effects of reconstruction algorithms, including prescribed artifact corrections and postprocessing methods. Image domain metrics typically involve standard IQ metrics [e.g., contrast-to-noise ratio (CNR) and noise power spectrum (NPS)], observer perceived image quality metrics (e.g., Likert score), and task-specific metrics (e.g., Dice coefficient in downstream segmentation tasks). We focus on projection domain metrics in this work (as described in Sec. II.D) and leave image domain metrics to a different study[24].

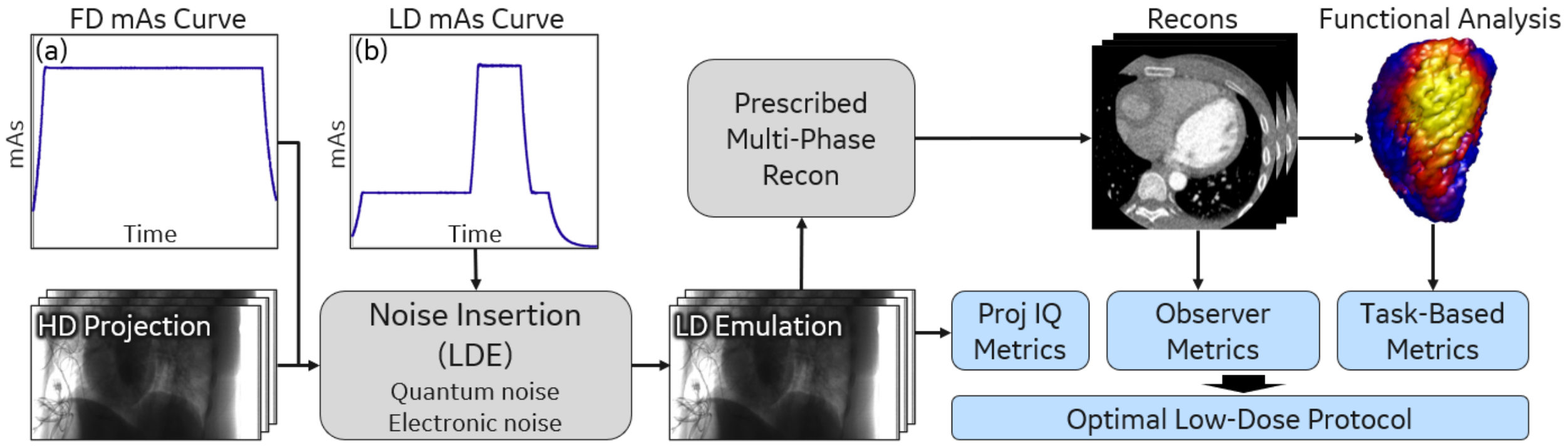

Figure 1. Workflow of the proposed low dose emulation (LDE) method and its targeted applications. The full dose (FD) data [mAs vs. time curve shown in (a)] is acquired as part of the current clinical workflow and provides the basis projection data for creating low-dose (LD) emulations. The noise injection process creates accurate emulated LD projections for any prescribed low dose mAs curves (b). By analyzing these emulated LD datasets in projection (focus of this work) and/or image domains (described in Ref [24]), minimum dose CT protocols can then be determined.

## II.A. Raw Data-Based Noise Emulation

For each view (view index denoted as $i$) within the complete cardiac scan window, the proposed LDE approach modifies the original FD CT projection data in a manner that simulates low mAs X-ray tube output. When the X-ray output is reduced by a factor of $r_i$ ($r_i < 1$ for every view), the mean signal is reduced by a factor of $r_i$, and additional noise needs to be added to emulate the corresponding signal-to-noise ratio (SNR) reduction. Following the previous work from our group,[12] we use a projection domain noise emulation framework where additional noise is modeled as two separate components: (1) Poisson distributed X-ray (i.e., quantum) noise in the *photon* domain ($n_{i,q}$), and (ii) Gaussian distributed electronic noise in the *electron* domain ($n_{i,e}$). These two components are uncorrelated as they are added in different domains. The overall LDE process is expressed as:

$$y_{i,LDE} = T\left(r_i \frac{y_{i,\mathrm{FD}}}{T} + n_{i,q}\right) + n_{i,e} \quad (1)$$

where $y_{i,FD}$ is the original full RR dose projection in electron domain at view $i$, $y_{i,LDE}$ is the low-dose emulated projection in electron domain, $T$ is the conversion factor from X-ray photon domain to electron domain, $r_i$ is again the X-ray output reduction factor for view $i$, which can be calculated by dividing the targeted LD mAs curve with the original FD mAs curve at each view [example mAs curves shown in Fig. 1(a) and (b)]. The injected quantum noise $n_{i,q}$ and electronic noise $n_{i,e}$ are assumed to be independent of $y_{i,FD}$. Since $y_{i,FD}$ is in the electron domain, it was first converted to the photon domain by dividing by $T$, scaled for the desired mAs level, contaminated with quantum noise in the photon domain, and then converted back to the electron domain. Finally, additional electronic noise was added in the electron domain.

Variance of the injected noise can be calculated as[12]:

$$\sigma_{i,q}^2 = (r_i - r_i^2)\sigma_{i,q,FD}^2 = (r_i - r_i^2)\frac{y_{i,FD}}{T} \quad (2)$$

$$\sigma_{i,e}^2 = (1 - r_i^2)\sigma_{e,sys}^2 \quad (3)$$

where $\sigma_{i,q}^2$ is the variance of the injected quantum noise in photon domain (i.e., variance of $n_{i,q}$), $\sigma_{i,q,FD}^2$ is the variance of the quantum noise of $y_{i,FD}$ in photon domain. Considering that X-ray photon detection is a Poisson process, $\sigma_{i,q,FD}^2$ is equal to the mean photon count $\frac{\overline{y_{i,FD}}}{T}$, which can be approximated with $\frac{y_{i,FD}}{T}$. Equation (3) shows the variance of the injected electronic noise in electron domain $\sigma_{i,e}^2$ (i.e., variance of $n_{i,e}$), where $\sigma_{e,sys}^2$ is the variance of the electronic noise of the CT system, which is assumed to be independent of the signal magnitude. We modeled $\sigma_{e,sys}^2$ as detector channel-dependent in this work.

To add the correct amount of X-ray noise in the photon domain, we need to model the energy integrating detection process which can provide a correct mapping between the photon domain and the electron domain for a given CT scanner. In this work, we approximate this mapping as:

$$y = D_G \sum_k E_k N_k \quad (4)$$

where $y$ is the measurement in the electron domain for each channel, $D_G$ is a detector gain factor denoting per keV of photon energy used to covert to electrons, $E_k$ is the mean energy of photons in the k$^{th}$ energy bin, and $N_k$ is number of detected photons at energy $E_k$. We can further simplify this transform by introducing $E_{eff}$, an effective energy satisfying $\sum_k E_k N_k = E_{eff} N$, where $N$ is total number of detected photons (i.e., measurement in photon domain). Then we have:

$$y = D_G E_{eff} N = TN \quad (5)$$

where the effective energy $E_{eff}$ can be approximated using the known X-ray spectrum, bowtie filtration, and beam-hardening corrected post-log measurement.[15]

In the LDE process, two system parameters need to be measured and calibrated: (1) $\sigma_{e,sys}$, the standard deviation of electronic noise; and (2) $D_G$, the gain factor per keV of photon energy. These two parameters are dependent on detector material variations and detector spectral dependency, the analog circuits (ASIC) used for detector readout, and many other properties of each detector channel, which make them vendor-dependent, scanner-dependent, and channel-dependent within the same scanner. We propose two experiment-based approaches to estimate the channel-dependent system parameters $D_G$ and $\sigma_{e,sys}$. Note that both $D_G$ and $\sigma_{e,sys}$ are not view-dependent (i.e., they are independent of $i$).

While the gain factor $D_G$ is channel-dependent, it is *not* assumed to be energy dependent. The separation of $D_G$ and the effective energy $E_{eff}$ is the key benefit of the approximation in Eq. (4). While $E_{eff}$ can be calculated analytically, $D_G$ requires experimental calibration as mentioned above. Because $D_G$ is energy-independent, this calibration is only required at one energy level. While this energy-independent assumption of $D_G$ is not perfect (due to physical effects such as depth-dependent light collection efficiency of the scintillator[25]), we find that it is accurate enough for our purposes as shown in Sec. II.C.

## II.B. Electronic Noise Modelling

The source of the electronic noise is the detection system built with analogue integrated circuits (ASIC) behind the detector scintillator; it produces background noise while the X-ray tube is turned on or off. The electronic noise is often modelled as a zero-mean Gaussian distribution assuming that the average dark current was subtracted. The variance of electronic noise [$\sigma_{e,sys}$ in Eq. (3)] can be estimated from a dark scan, i.e., an acquisition without X-ray exposure. We used a non-clinical scan mode of the GE HealthCare Revolution™ (Waukesha, WI) to acquire ~4000 views with full collimation (16cm) without X-rays.

Our empirical analysis of dark scan data indicated that the variance of the electronic noise is channel dependent due to channel dependent properties of the underlying ASIC (electronic noise level standard deviation: $\sigma$ = 1.93 ADU). Due to temporal correlation between consecutive views (i.e., memory effect of the ASIC), an independent Gaussian distribution is not the best model of the electronic noise. Therefore, instead of using a simple zero-mean Gaussian distribution with the same variance for all channels, we propose an empirical electronic noise realization approach utilizing the experimentally acquired dark scans.

The first step in our electronic noise generation is to randomly sample a series of views from a large database of measured dark-scan views as described above. While this ensures an extremely realistic channel-specific signal distribution, the random sampling unfortunately removes the native temporal correlation discussed above. As a result, we next restore the original temporal correlation by applying a symmetric 3-tap temporal (view-axis) filter. The coefficients of the filter are channel-specific and chosen to minimize the difference between the temporal autocorrelation of the original data for a given channel and that of the corresponding sampled-then-filtered data. There is only one controlling parameter $k$ for the 3-tap filter for each pixel (the three temporal coefficients for the filter are: $-k, 1+2k, k$ respectively). Across the entire detector, $k$ has a mean value of 0.132 and a standard deviation of 0.011. Through this process, channel-dependent variance and temporal correlation of the electronic noise are properly modeled.

## II.C. Channel-Dependent $D_G$ Estimation

As mentioned in Sec. II.A, if the electronic noise is ignored, the signal measured in the electron domain follows a Poisson distribution with channel-dependent scalar factor, $D_G E_{eff}$, where the ray-dependent $E_{eff}$ can be computed using a known X-ray spectrum, additional filtrations, and beam hardening corrected post-log views. However, $D_G$ is a detection system specific parameter which needs to be measured or estimated for each scanner and each detector channel. We propose an experiment-based approach to estimate the system parameter $D_G$, which follows the following three steps: (1) we scanned various thicknesses of aluminum (Al) plates at different dose levels by varying the X-ray tube current (mA); (2) we created mean vs. variance (M-V) curves for each detector channel from aluminum plate scans; and (3) following Eq. (1), we created emulated M-V curves with various $D_G$ values and found the $D_G$ that minimizes the difference between the emulated and the real M-V curves in step (2); Note that the system tested in this work has an mA lower limit. Therefore, Al plates are utilized to create M-V curves at even lower signal levels. Following the same procedure, a different scanner can be characterized to produce a new set of slopes of M-V curves, which will then be translated into the system parameter $D_G$. These calibration steps are only required for one energy level (as $D_G$ is *not* energy dependent). Details of each step follow.

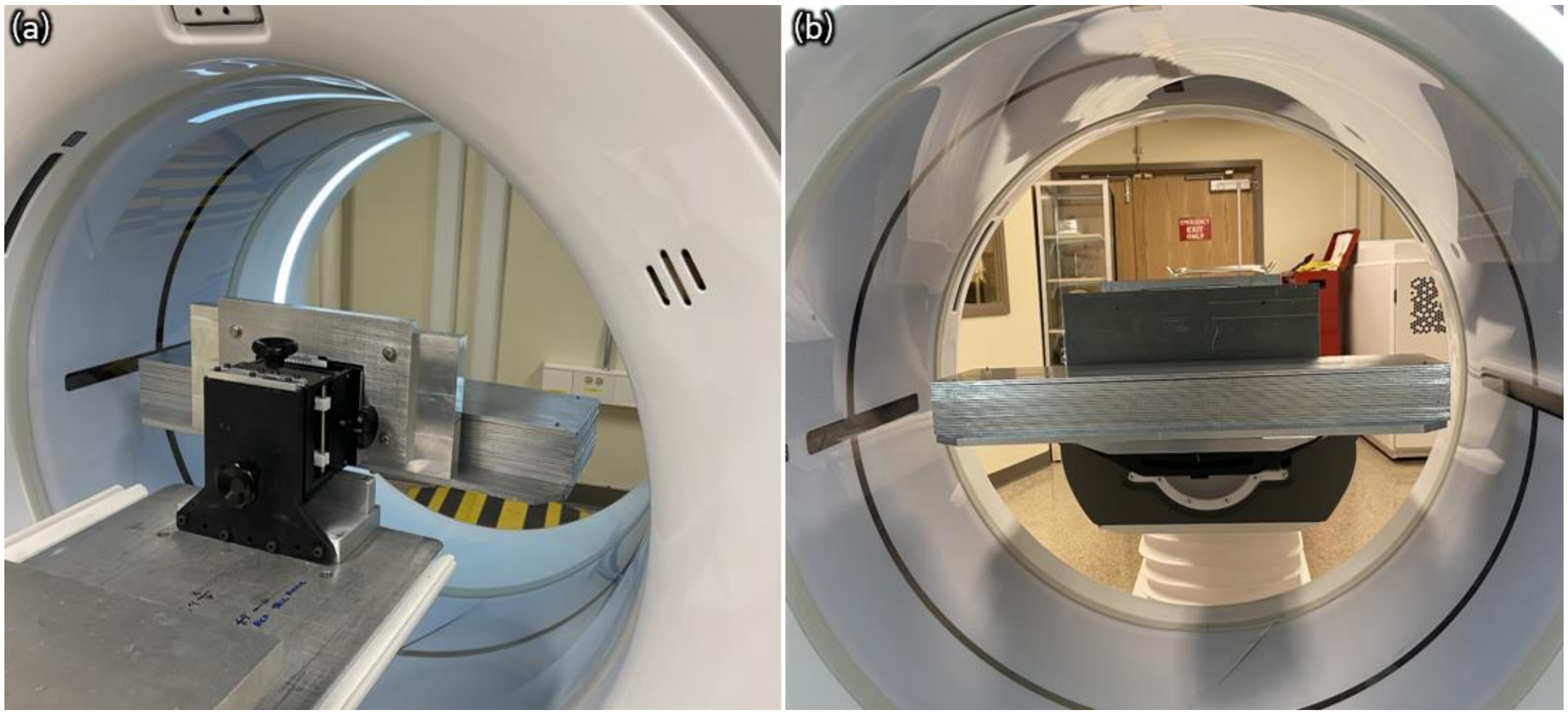

**Figure 2.** Experiment setup for stacked Al plates. The thickness of each plate is 1/8 inch. Stacked plates are cantilevered and raised to cover the detector channels during acquisition. The position of the X-ray tube is fixed at 12 o'clock [i.e., image top in (b)] during acquisitions.

In the first step, we scanned stacked Al plates of 1/8 inch thickness cantilevered to the bore of GE HealthCare Revolution scanner as shown in Fig. 2. We used the non-clinical scan mode to fix the location of the X-ray tube at the top of the gantry and acquired ~ 4000 views without rotating the gantry. The stack of plates was positioned high enough to provide good detector coverage. At each acquisition, the plates are either removed or added to create different thickness measurement in the range of 3/8 inch through 23/8 inch. The experiments were repeated with three tube currents (10, 25, 600 mA). To validate our assumption that $D_G$ can be modeled as an energy-independent parameter with minimal loss in accuracy, two tube voltages (80 kVp and 140 kVp) with and without bowtie filtration are used in the calibration step. In practice, $D_G$ calibration is only required at one energy level as mentioned above.

Next, we computed channel means and variances from the measurements of varying Al plates thickness and created one Mean-Variance (M-V) curve per detector channel. Since the signal measured in the electron domain mostly follows a Poisson distribution with channel-dependent scalar factor, $D_G E_{eff}$, we expect M-V curves to be mostly linear and their slopes to reflect channel-dependent scalar factors, $D_G E_{eff}$. Therefore, slopes of M-V curves become larger for scanning at higher kVp or thicker objects where $E_{eff}$ is higher due to acquisition voltage or beam-hardening as shown in Fig. 3(a). Figure 3(c) shows the same set of curves as Fig. 3(a) with their variance values divided by $E_{eff}$. This way, slopes of these new curves are determined by $D_G$ only. The close correspondence between the curves at 80 and 140 kVp (at least in the linear portion) supports our assumption that $D_G$ can be modeled as an energy-independent parameter. This allows us to perform calibration at only one energy level and applies the results to other targeted energy levels by changing only the analytical $E_{eff}$ term.

At extreme low signal / flux level (i.e., very low channel mean), the system exhibits M-V nonlinearity as shown in Fig. 3(b). This type of system nonlinearity has also been observed by other researchers (e.g., Ref [8]). This is due to built-in signal-dependent processing from the CT system that is designed to avoid negatives when taking the log-transform during the reconstruction stage when the signal is very low. This type of signal dependent processing will typically reduce the variance of the signal as can be observed in Fig. 3(b). In our experiment, we observed that this effect will be diminished (i.e., curves become more linear) when the channel mean is equal or larger than around 100 detector readout counts, which can be translated into $3.3*\sigma_e$, or ~ 8 photons at 80kVp acquisition.

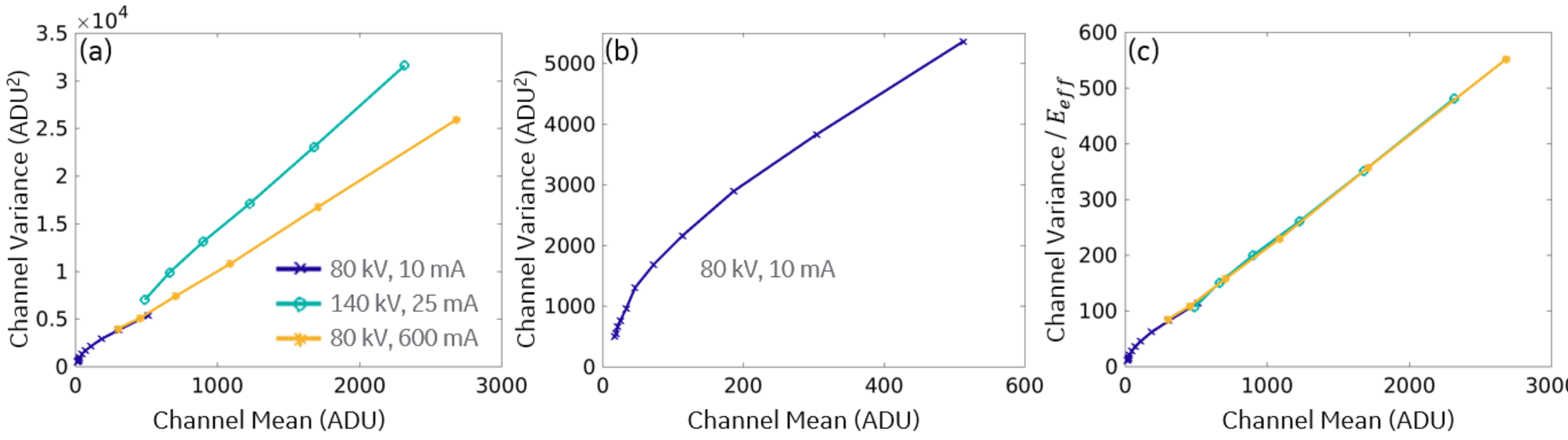

**Figure 3.** Sample mean-variance (M-V) curves from one example detector channel at the detector center. The units of the channel mean is in detector readout count (analog-to-digital unit, ADU). Each data point corresponds to one measurement. M-V curves are mostly linear unless the channel mean is very small; for example, the nonlinearity at extremely low flux levels is observed at 80kVp, 10mA, shown in the zoomed view (b). The same curves in (a) are also shown in (c) with their variance values divided by their corresponding effective energies ($E_{eff}$), confirming that $D_G$ [slopes in (c)] is independent of energy levels.

In the third step, we ran noise emulation with various $D_G$ values and created emulated M-V curves for every detector channel. Then we computed the root mean square error (RMSE) between emulated M-V curves and acquired M-V curves as a function of $D_G$. By fitting a quadratic function (second order polynomial) to the RMSE curves, we picked $D_G$ yielding the best match between the LDE results and the LD acquisition per channel. This step is necessary due to the presence of non-linear effects noted above (otherwise one can directly calculate $D_G$ by dividing slopes of the M-V curves by $E_{eff}$).

## II.D. Low Dose Protocol Optimization

The goal of protocol dose optimization is to deliver just enough dose to each patient by selecting the optimal tube voltage and current for each patient size while achieving desired image quality for a given downstream task. Instead of using traditional image domain IQ metrics such as CNR and NPS, we propose a raw data-based image quality metric indicating how many projection rays fall under an extreme low flux condition thus introducing undesirable effects on preprocessing and reconstruction steps.

When X-ray flux is below 3.3*$\sigma_e$, or below 6-8 photons per channel (depending on x-ray energy), the measured signal no longer strictly follows a Poisson distribution. As noted in Sec. II.C, nonlinearities in M-V curves can be observed in this extreme low flux region. Very low signal levels initiate custom designed signal-dependent processing which nonlinearly transforms the noise texture. When many projection rays are under this extreme low flux condition, the resulting images can be significantly corrupted with dark shading, and bright / dark ring artifacts. Motivated by this effect, we introduce two thresholds: (i) an extreme low flux threshold (ELF threshold), and (ii) an image quality threshold (IQ threshold, %), to define a critical flux level below which the image quality is not suitable for 4D Cardiac CT post processing (e.g., segmentation or motion correction).

First, if a detector channel signal level is smaller than 3.3*$\sigma_e$, we define this channel to be under an extreme low flux condition (ELF condition). Second, the proposed IQ metric denotes the percentage of total projection rays that are *above* the ELF condition across the detector over the entire acquisition (the higher the better). From our empirical experience, CT preprocessing and reconstruction steps can produce reasonable images even with only 90% of projection rays above the ELF condition. In this case, the IQ threshold is 90%. IQ threshold can vary among clinical applications; for simple LV segmentation to compute ejection fraction, the IQ threshold can be quite low (even ~85%).

While the ELF threshold (e.g., 3.3*$\sigma_e$) used in this work is determined for the GE HealthCare Revolution system, corresponding thresholds for other systems can be determined by finding the readout count threshold where strong non-linear effects occur, which can be observed with the M-V curves when doing the $D_G$ calibration step in Sec. II.C [Fig. 3(b)].

Finally, we defined a "critical mA" level (CR mA) as the lowest tube current which meets the IQ threshold. We used the proposed LDE framework with the calibrated system parameters to generate emulated projections at many mA levels for each kVp setting. Furthermore, the X-ray dose achieved by scanning with the critical mA can be computed using a size specific dose estimate (SSDE) factor[26] of the scanned object and Monte-Carlo simulation.[15] Since different kVp settings will provide different critical mA values for the same object, protocol optimization can be performed by choosing kVp value of the lowest SSDE adjusted $CTDI_{vol}$ and the critical mA value for that kVp setting, as will be shown in Fig. 9 below.

## II.E. Experiments

Phantom and clinical studies were performed using a GE HealthCare Revolution$^{TM}$ (Waukesha, WI) scanner as shown in Fig. 2. For phantom studies: we scanned the CIRS Tissue Equivalent abdomen phantom (lateral size of 42 cm) (CIRS, Norfolk, VA) at 80kVp with two different mA levels: 5.6 mAs (LD) and 168 mAs (FD). We emulated the LD acquisition from the FD acquisition and compared the emulated acquisitions with measured LD acquisitions. Filtered backprojection (FBP) reconstructions with the standard kernel (built into the scanner console) were used to reconstruct all images in this work.

The clinical study was approved by the institutional review board (IRB) with a waiver of informed consent. The biggest challenge in validating our LDE approach with clinical scans is that these scans typically don't have matched FD and LD acquisition pairs. To address this issue, we retrospectively obtained an acquisition involving mA modulated cardiac scans which contained multiple rotations with low mAs (LD) except at systolic and/or diastolic phases, as shown in Figure 4. We used FD (red box, 168 mAs) projection to generate a LDE projection at 56 mAs and compared it with two measured LD acquisitions at 56 mAs, which are 2.5 rotations (green box) and 2 rotations away (orange box) from the FD acquisition respectively. While the FD and the two LD acquisitions are not exactly matched in terms of the underlying anatomy due to heart motion (not the same phase), their mean signal values are relatively similar at each detector channel considering that their ray or conjugate ray angles are matched (separated by $N/2$ rotations, where $N$ is an integer number).

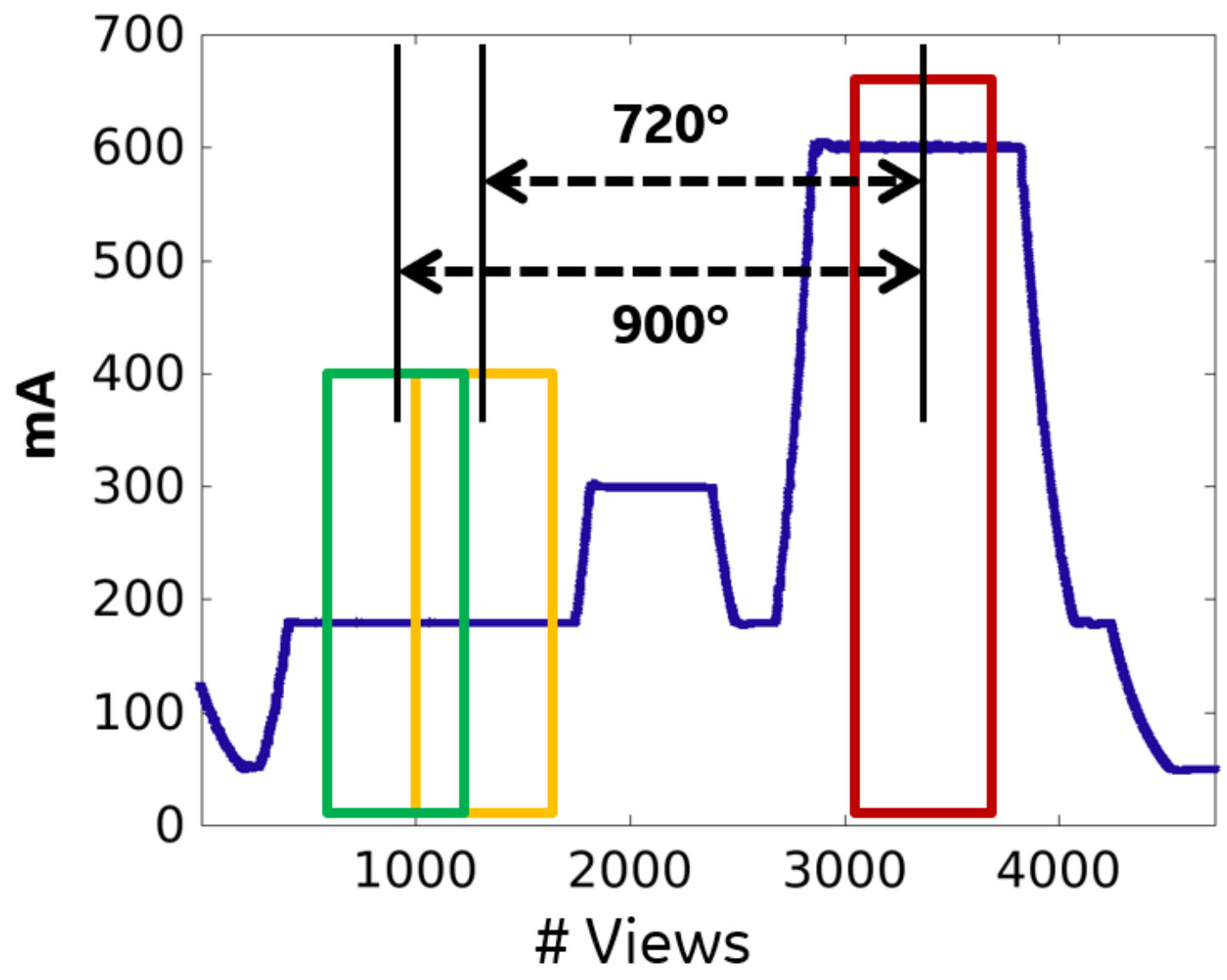

**Figure 4.** Tube current (mA) profile of the FD (full-dose) scans used for LD (low-dose) emulation, where we have FD data at the diastolic phase (red box), and LD data that are 2 rotations (yellow box) and 2.5 rotations (green box) from the FD data. Note that FD and LD data have the same center ray angle.

# III. Results

## III.A. Spatially Varying $D_G$

Figure 5 shows the performance of the proposed LDE method with three different detector gain ($D_G$) settings: (i) channel-independent $D_G$ optimized on GE HealthCare Discovery™ CT750 HD system from our previous work;[12] (denoted "original $D_G$") (ii) channel-independent $D_G$ optimized on the GE HealthCare Revolution system used in this work (denoted "globally optimized $D_G$"); (iii) channel-dependent $D_G$ optimized on the GE HealthCare Revolution system (denoted "channel-dependent $D_G$"). M-V curves from three detector channels at various locations are reported.

With the original $D_G$ setting, M-V curves from the measured LD acquisitions and the emulations show visible discrepancies. This suggests that $D_G$ from other scanners will not work well for noise emulation at very low flux. With estimated $D_G$ for the targeted scanner (globally optimized $D_G$ and channel-dependent $D_G$), we get a better match between M-V curves from the measured LD acquisitions and the emulations. As shown in Figure 5(b) and (c), M-V curves from different detector channels show varying slopes and deflection points, indicating channel-to-channel variation in $D_G$. Channel-dependent $D_G$ estimation results in the closest match between the measured LD acquisitions and the emulations as shown in Figure 5(c). This emphasizes the importance of channel-dependent $D_G$ calibration for the targeted scanner. For quantitative comparison, RMSE of the emulated and real (experimentally acquired) M-V curves are plotted for each pixel as shown in Fig. 5(d-f). The proposed method resulted in the lowest RMSE as shown in Fig. 5(f) due to the usage of channel-dependent $D_G$ estimations. One can see that RMSE shows a clear channel-dependent pattern in Fig. 5(e), which was mostly removed with the proposed method, indicating more accurate noise emulation for all pixels.

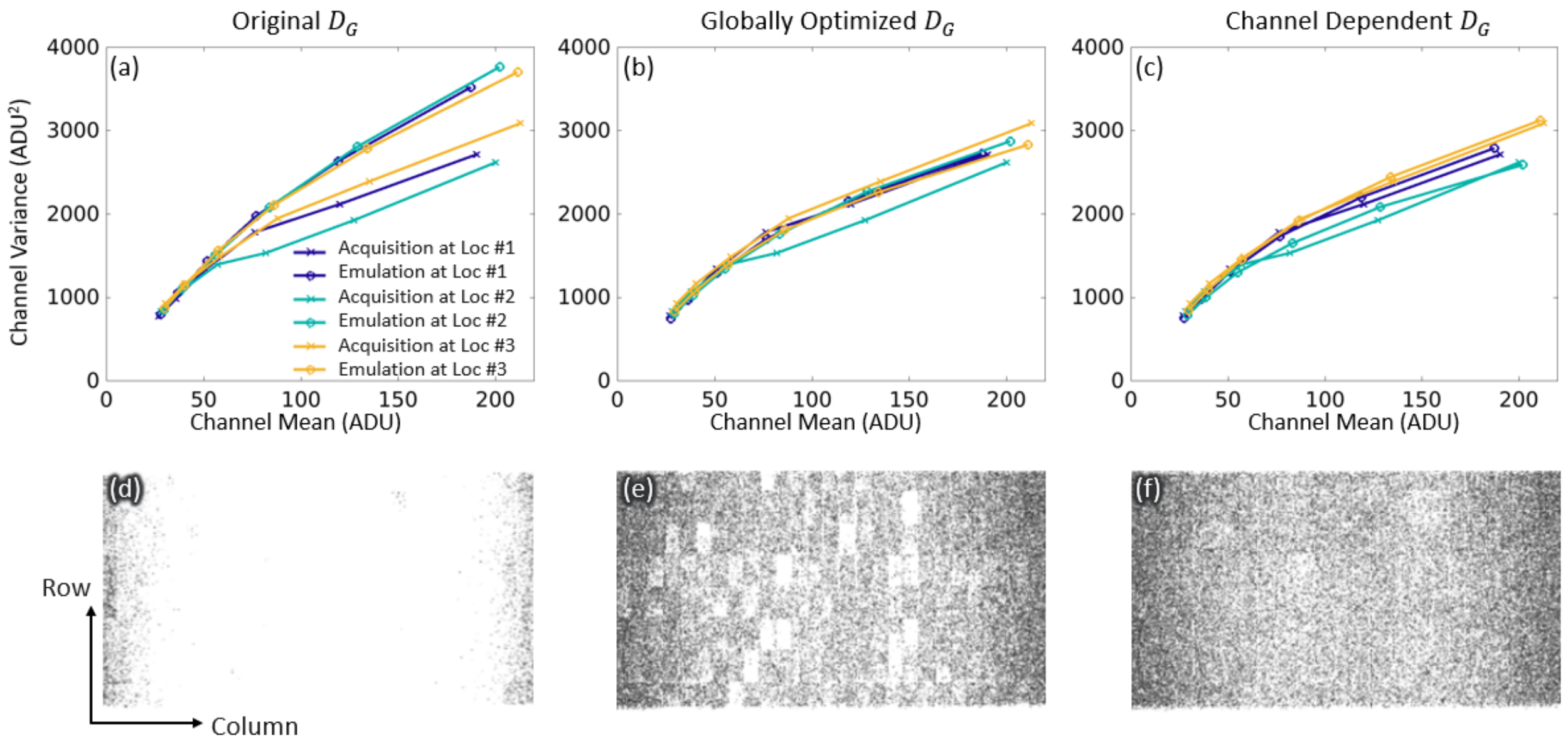

**Figure 5.** M-V curves from three detector channels before and after dector gain ($D_G$) optimization: (a) with original $D_G$, (b) with globally optimized $D_G$, (c) with channel-dependent $D_G$. Each colored line represents an independent channel location (the three locations are column 416, 583, 749 respectively at the central row of the detector, i.e., from center gradually going to the edge). Lines with circles correspond to M-V curves from emulation and lines with crosses correspond to M-V curves from acquisition. M-V curves from the emulations align best with M-V curves from the acquisitions with channel-dependent $D_G$. RMSE in terms of acquisition and emulation M-V curves for each pixel are plotted for (d) with original $D_G$, (e) with globally optimized $D_G$, (f) with channel-dependent $D_G$, respectively. Edge of the detector were cropped out because of the bowtie filter. Display window: [0 500] $ADU^2$.

## III.B. Phantom Studies

We further validated the proposed LDE framework with phantom studies (detailed in Sec. II. E) using the channel-dependent $D_G$ setting reported above. The images from the emulation closely recreated low flux related artifacts, such as dark shading around objects, dark and bright rings near center of field of view, and channel dependent artifacts in sagittal slices as shown in Fig. 6. The residual differences between the emulated and measured LD acquisition are likely related to the signal nonlinearity when the flux level is below the ELF condition. Figure 6 also shows radially averaged axial NPS for measured and LDE images (RMSE=6.85 $HU^2/mm^2$), demonstrating strong agreement between the measured and emulated LD images in terms of the noise magnitude and noise spectrum / texture.

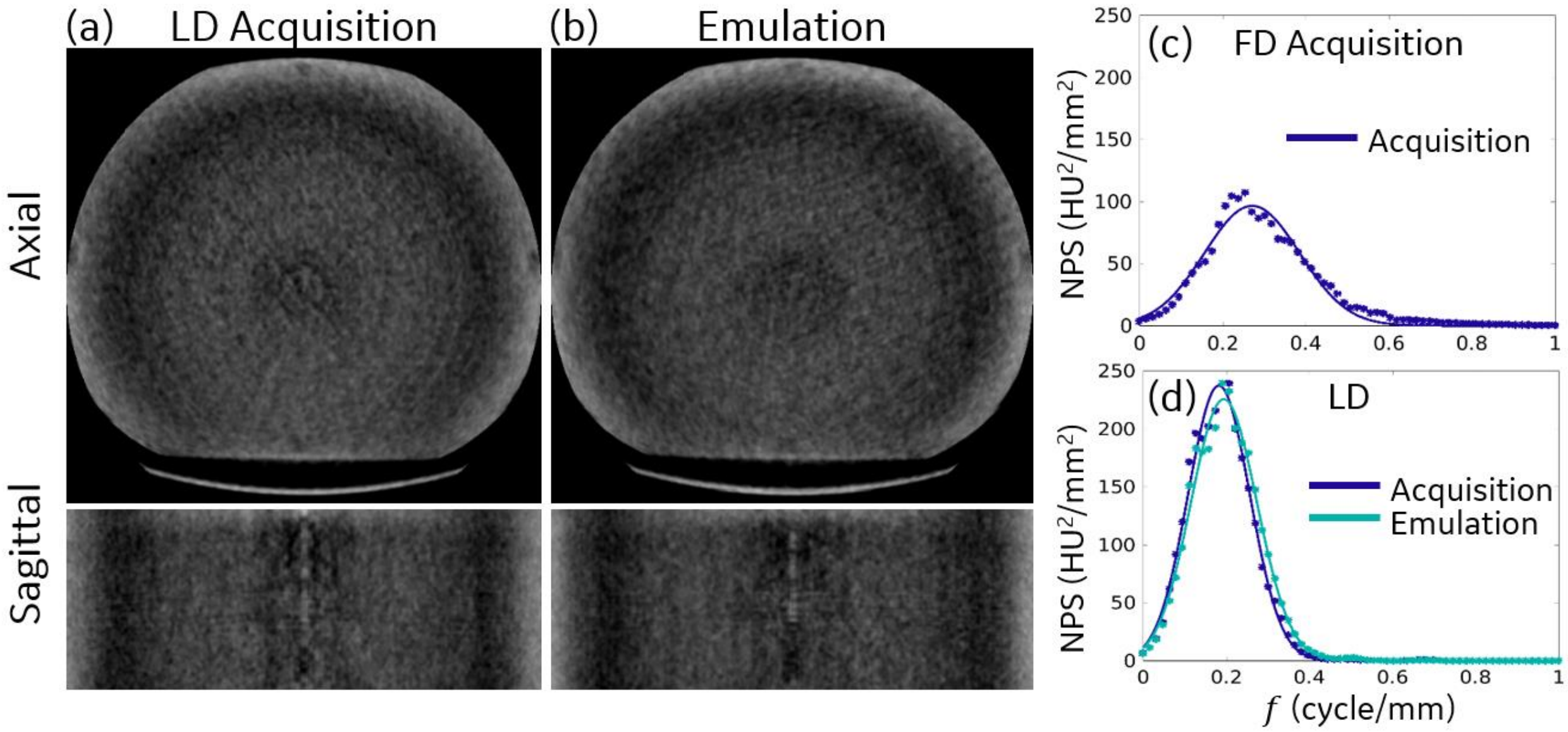

**Figure 6.** Reconstructed images of the 42cm CIRS abdomen phantom in axial and sagittal planes: (a) from the measured LD acquisition and (b) from the LDE emulated projection. The phantom was scanned at 80 kVp and 5.6 mAs where only 27.8% of rays are above the ELF condition. Images are displayed at [1000, 1200] (center, width) HU. Low flux related artifacts in the LD acquisition, such as dark shading patterns, and dark and bright rings near the center of field of view, were recreated in the emulation. Noise power spectrum were measured in the axial plane: (c) NPS for measured FD acquisition. (d) NPS for measured and emulated LD acquisitions at 5.6 mAs.

## III.C. Clinical Studies

In addition to the phantom study, we tested the proposed LDE approach with clinical cardiac studies as detailed in Sec. II. E. Figure 7 shows a side-by-side comparison of the two measured LD images and the LDE images. While the FD acquisition used for emulation came from the diastolic phase, the measured LD acquisitions came from other cardiac phases as explained above. Therefore, the anatomical features in images from the emulation and the measured acquisitions, shown in Fig. 7(c) and Fig. 7(a)(b), don't exactly match. However, qualitatively, the level and texture of noise in the LDE image show a close resemblance to those in the measured LD images. Quantitatively, the LDE image shows realistic CNR between the left ventricle blood pool and myocardium. The average CNR disagreement between measured LD and LDE images are less than 1.1%. We note that while CNR is a commonly used metric for quantifying the noise, it is insufficient to fully reflect the perceived image quality difference between emulated and real LD images, due to nonlinear low signal correction in the reconstruction process. These corrections often involve data smoothing which trades spatial resolution for noise reduction. Additionally, the discrepancy of HU values [blue ROIs in Fig. 7(d)] and resolution (width of the edge spread function[27] of the myocardium blood pool) between the emulated and the real LD images are less than 1.5%.

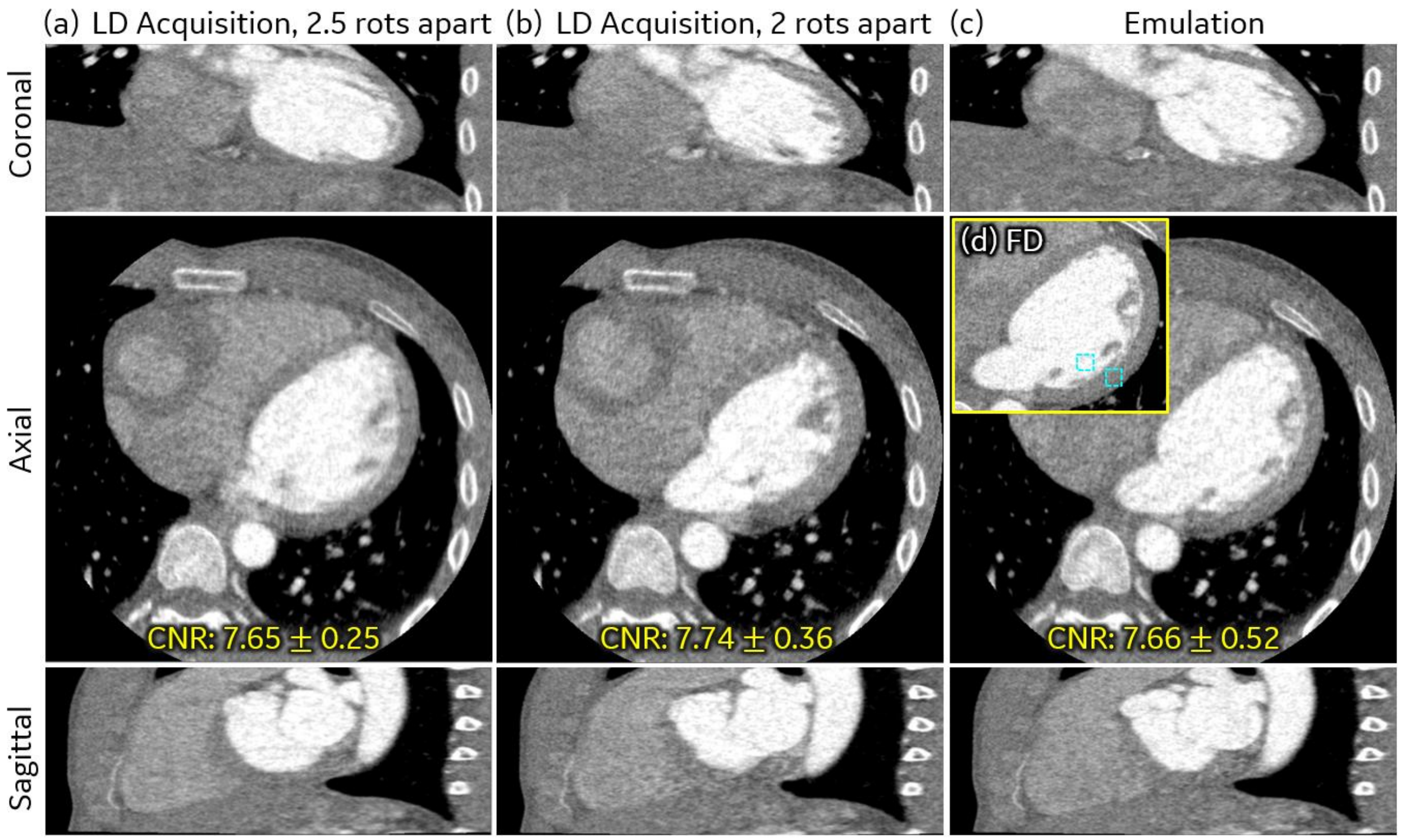

**Figure 7.** Comparison of image quality from LD acquisitions and the emulated LD images. (a) images from the LD (low-dose) acquisition that is 2.5 rotations apart from the FD (full-dose) acquisition, (b) images from the LD acquisition that is 2 rotations from the FD acquisition, (c) emulated images from the LDE projection. (a) and (b) were acquired at 56 mAs and (c) was emulated at 56 mAs, where the IQ metric is 95.5%. Note that the cardiac phase of the LDE projection is different from those of LD acquisitions. Images are displayed at [1120, 1200] (center, width) HU. Cyan ROIs in subfigure (d) was used to measure CNR for LD acquisitions and the emulation.

## III.D. Optimized Low Dose Protocol

Motivated by realistic noise emulation results using the proposed LDE approach, we used the proposed projection domain IQ metric for protocol optimization in 4D Cardiac CT, which requires acquisitions of objects of different sizes at various mA and kVp settings.

To calculate the critical mA for objects of different sizes, CIRS Tissue Equivalent phantoms with 5 sizes were each scanned at the maximum possible mA at five different kVp levels (70 kVp, 80 kVp, 100 kVp, 120 kVp, and 140 kVp) while keeping the gantry rotation speed at 0.28 sec/rot. From these 25 full dose scans, we then emulated LD scans at mA levels shown in Fig. 8. At each mA level, we computed the projection domain IQ metric (i.e., percentage of rays above the ELF condition) for each phantom and for each kVp level, as shown in Fig. 8. As expected, low kVp acquisitions require higher mA to achieve the same IQ threshold. In this example, the critical mA of the 95% IQ threshold is marked with pink dashed lines. At the 0.28s gantry period, the 95% IQ threshold is not achievable at 70 and 80 kVp for the very largest phantom [see Figure 8(c)].

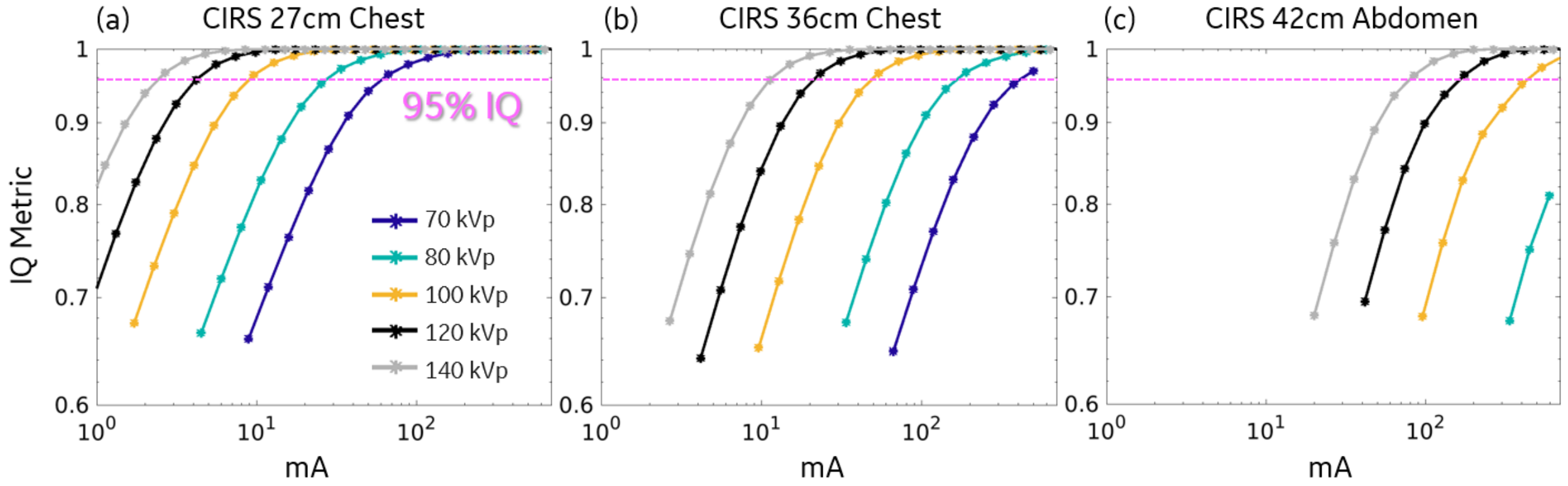

**Figure 8.** Critical mA plots at different kVp settings for objects of different sizes: (a) CIRS 27 cm chest phantom, (b) CIRS 36 cm chest phantom and (c) CIRS 42 cm abdomen phantom. Each colored line represents different kVp acquisitions. Their y axes denote the IQ metric. For a given phantom size and kVp, as mA increases, the IQ metric increases. The pink dashed lines mark the 95% IQ threshold.

We computed the SSDE adjusted $CTDI_{vol}$ for the five kVp levels for each CIRS chest phantom using the critical mA values. The dose computation was performed using the CatSim toolkit.[15] Note that throughout this section, the SSDE adjusted $CTDI_{vol}$ is reported for a full heartbeat 4D scan (assuming 60 bpm). We plotted 4D SSDE adjusted $CTDI_{vol}$ against effective diameter at different kVp levels. The scanner has a manufacture-specified minimum 50 mA threshold in cardiac mode to ensure adequate image quality, which explains the rise of SSDE adjusted $CTDI_{vol}$ measurement on the left side of the plots. Two IQ threshold levels (defined in Sec. II.D), 85% and 90%, were considered as shown in Fig. 9.

With an IQ metric threshold of 85%, 70 kVp provides the lowest dose for DE below 28 cm and 140 kVp provides the lowest dose for DE above 43 cm as shown in Figure 9(b). By selecting the optimal kVp for a given size and using corresponding critical mA, the 4D SSDE adjusted $CTDI_{vol}$ of the optimized protocol stays below 2.1 mGy for DE below 35 cm and 3.6 mGy for all clinically relevant DE. Utilizing this approach, we can find the optimized protocol for any effective diameter at any given IQ threshold. The recommended protocol for 4D functional cardiac imaging is shown in Table 1.

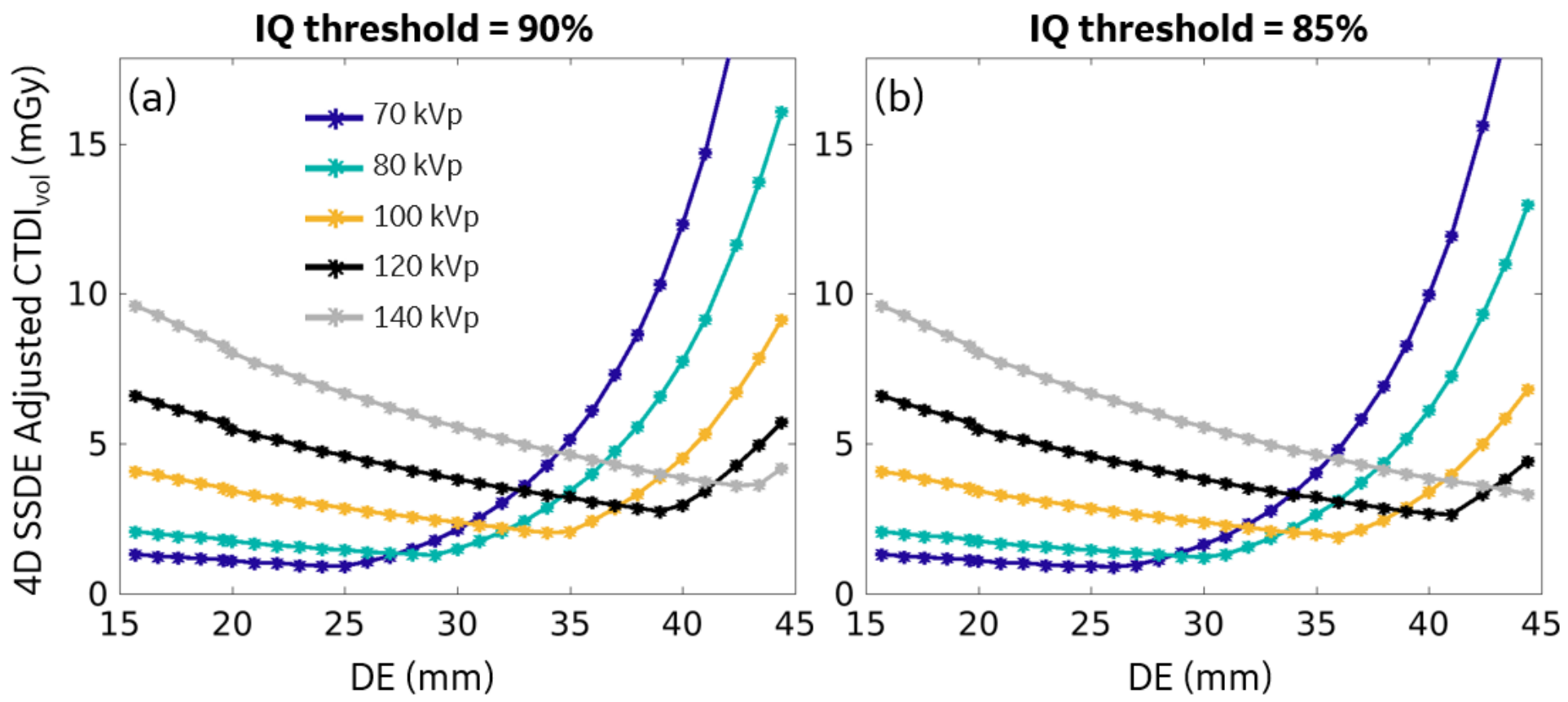

**Figure 9.** Dose vs. Patient size (chest region) curves: (a) for IQ threshold of 90%, and (b) for IQ thresholds of 85%. For a given acquisition kVp, SSDE adjusted $CTDI_{vol}$ at the critical mA for IQ threshold of 90% and 85% were computed for a wide range of effective diameters. Note that the dose reported here is for a full heartbeat 4D scan (assuming 60 bpm). Minimum system allowable mA was set to 50mA. DE = effective diameter

| Effective diameter (cm) | 4D Cardiac CT Protocol (SSDE adjusted $CTDI_{vol}$) for given IQ threshold with minimum 50 mA limit | | | | |
|---|---|---|---|---|---|
| | **95%** | **90%** | **85%** | **80%** | **75%** |
| 20 | 70kVp, 50mA (1.11mGy) | 70kVp, 50mA (1.11mGy) | 70kVp, 50mA (1.11mGy) | 70kVp, 50mA (1.11mGy) | 70kVp, 50mA (1.11mGy) |
| 25 | 70kVp, 74.4mA (1.35mGy) | 70kVp, 50mA (1.35mGy) | 70kVp, 50mA (0.92mGy) | 70kVp, 50mA (0.92mGy) | 70kVp, 50mA (0.92mGy) |
| 30 | 80kVp, 90.3mA (2.18mGy) | 80kVp, 61.8mA (1.50mGy) | 80kVp, 50mA (1.21mGy) | 80kVp, 50mA (1.21mGy) | 70kVp, 64.9mA (1.00mGy) |
| 35 | 100kVP, 79.6mA (3.14mGy) | 100kVp, 52.4mA (2.07mGy) | 100kVp, 50mA (2.00mGy) | 100kVp, 50mA (2.00mGy) | 80kVp, 81.6mA (1.60mGy) |
| 40 | 140kVp, 50mA, (3.85mGy) | 120kVp, 55mA (2.96mGy) | 120kVp, 50mA (2.69mGy) | 120kVp, 50mA (2.69mGy) | 100kVp, 65.8mA (2.18mGy) |

**Table 1.** Recommended kVp and mA settings for low dose 4D Cardiac CT acquisitions (covering the full cardiac cycle, assuming 60 bpm) targeting different IQ thresholds (the higher the better).

## IV. Discussion and Conclusions

In this work, a new projection domain low dose emulation approach was developed to emulate realistic low dose 4D cardiac scans from existing full dose cardiac scans. One possible application of the proposed method is for functional cardiac CT protocol optimization (kV and mA, as shown in Table 1), i.e., determining the lowest possible patient dose required to produce images with sufficient quality for the targeted downstream clinical tasks such as measuring ejection fraction, myocardial strain, etc.

We proposed channel-dependent system parameter estimation to fine-tune noise emulation performance at very low x-ray flux. The variations in electronic noise statistics and gain factor $D_G$ among detector channels likely originate from variations in the detector scintillator and in underlying readout circuits of each channel. If we are targeting noise emulation where detector readout counts of projection rays are larger than 3.3*$\sigma_e$, the improvement provided by channel-dependent electronic noise modeling starts to diminish because electronic noise variation between channels is overwhelmed by X-ray noise statistics. Also, the variation in detector gain ($D_G$) between different scanner generations was observed to be larger than the variations among detector channels in the same scanner. The standard deviation of $D_G$ across different detector channels of the Revolution system is ~6.5% while the difference between $D_G$ in the Revolution and Discovery systems is ~28%. Therefore, the proposed empirical characterization of system parameters significantly improves the quality of noise emulation when different models of scanners are used. Channel-dependent $D_G$ will provide additional improvement only when noise emulation targets very low flux acquisition as shown in Fig. 5.

Physics effects, such as channel-specific characterization of quantum and electronic noise, beam hardening effects, nonlinearity in detector response at low flux conditions, and temporal correlation in electronic noise, were included in the LDE approach. However, there are still some theoretical limitations: (i) The compound Poisson X-ray attenuation noise model was replaced with the Poisson only model. However, our experience with this simulation tool suggests that this approximation is negligible; (ii) Spatial correlation of the quantum noise (e.g., from detector x-ray fluorescence crosstalk) was not modelled; (iii) The gain $D_G$ was assumed to be energy-independent [i.e., separated from $E_{eff}$ as shown in Eq. (5)]. While this greatly simplifies the calibration process (no need for repeated calibration for each kVp), it could potentially reduce emulation accuracy due to depth-dependent light collection efficiency of the scintillator. This approximation was demonstrated to provide reasonable emulation results as shown in Fig. 5(c).

While this work focuses on mA modulation as a function of cardiac phase, the proposed LDE approach can also be used to emulate mA modulation as a function of gantry angle. This is typically used as part of the automatic exposure control (AEC) module, where higher mA was prescribed when X-rays were passing through the long axis of the elliptical objects or denser parts of the patients.

Utilizing the aforementioned LDE approach, a global projection domain IQ metric (percentage of rays above the extreme low flux (ELF) condition) was proposed to efficiently perform protocol optimization. Note that this IQ metric was not used to evaluate the accuracy of noise injection, but to perform simple procotol recommendations. Recommended protocols in terms of kVp and mA were then calculated based on an effective diameter (DE) of patients, which can be computed by measuring post-anterior (PA) length and lateral (LAT) length from previously reconstructed images or scout scans with PA and LAT views. This metric lets us calculate the minimal cost of 4D functional cardiac CT in terms of added dose relative to the standard 3D cardiac CT. An mA-modulated 4DCT scan (i.e., same mA as the standard 3D CT scan at the targeted phase, low-dose at any other phases) could provide temporal visualization of cardiac anatomy/functional changes over the full heart cycle. Taking an average patient size (DE) of 30 cm and an average heart rate (HR) of 60 bpm, such mA-modulated 4DCT scan targeting a 90% IQ threshold has only 63% higher dose than the minimum mA modulation width 3D CT scan (with an institutional average mA of 500). At the same time, the 4DCT x-ray exposure time is ~400% that of the standard 3D CT scan, allowing reconstruction of every phase in the cardiac cycle rather than a single phase. Note that these recommended protocols are only for cardiac CT (chest region) and need to be recomputed for other targeted anatomy due to difference in tissue material.

The proposed projection domain IQ metric is a straightforward metric that only focuses on the percentage of projection rays above the ELF condition. While this metric directly correlates with various low-dose related artifacts (these artifacts are typically related with rays below the ELF condition), it does not take into accounts factors like patients of different nominal sizes (e.g., pediatric vs. adult patients) or organs of interest. Images with the same IQ metric can look very different (e.g., rays below the ELF condition from heavy metal will likely create more artifacts than that from dense tissue). More advanced projection domain metric can certainly be envisioned in future work, for example: (i) metrics masked to the targeted organ; or (ii) metrics normalized to specific patient groups. Nevertheless, the proposed metric directly captures the number of X-rays that are causing artifacts (rays below the ELF condition) and is effective to a large extent.

Additionally, while a projection domain IQ metric is the focus of this work due to its simplicity, task-specific image domain IQ metrics are also the subject of our other work.[24] The optimized protocol can be quite different based on downstream functional analysis tasks in cardiac CT. For example, global functional metrics such as global longitudinal shortening (GLS) or ejection fraction are much more robust to noise compared to local metrics such as directional myocardium wall thickening. Task-specific image domain IQ metrics are likely necessary to provide more accurate protocol optimization results in these scenarios.

In summary, detailed steps to estimate system-dependent parameters for the raw-data based LDE framework were described, which was demonstrated to reliably produce realistic noise levels and texture with phantom and clinical studies. One potential application of the LDE approach is explored in this work, i.e., for protocol optimization to deliver the lowest dose while providing proper image quality for 4D cardiac CT acquisitions. Another potential application is to generate matched pairs of LD and FD data for training accurate deep learning-based denoising approaches,[28–30] which will likely benefit from the increased noise emulation accuracy of the proposed method. In addition, we are currently exploring the possibility of performing 4D Cardiac CT dyssynchrony analysis with dramatically lower radiation dose, which involves the proposed LDE approach and task-specific protocol optimization in image domain.

## Acknowledgement:

Research reported in this manuscript was supported by NHLBI of the National Institutes of Health under award number R01 HL14678 (PI: Dr. McVeigh). The content is solely the responsibility of the authors and does not necessarily represent the official views of the National Institutes of Health.

## Conflict-of-Interest:

Authors Z.Y, P.W, and J.P are employed with GE Healthcare. Author E.M has founder shares in Clearpoint Neuro Inc. and receives research funding from GE Healthcare, Abbott Medical, Pacesetter Inc.